\documentclass[a4paper, amsfonts, amssymb, amsmath, reprint, showkeys, nofootinbib, twoside]{revtex4-1}

\usepackage{graphics}      
\usepackage{graphicx}      
\usepackage{longtable}     
\usepackage{url}           
\usepackage{bm,color}            
\usepackage{amssymb, amsmath, enumerate, theorem, epsfig,subfigure,setspace}
\usepackage{amsfonts}

\newcommand{\nb}[1]{\textcolor{black}{#1}}
\newcommand{\nn}[1]{\textcolor{black}{#1}}

\begin{document}
\title{Managing Flow of Liquid Light}
\author{Nikita Stroev$^1$ and Natalia G. Berloff$^{1,2}$ }
\email[correspondence address: ]{N.G.Berloff@damtp.cam.ac.uk}
\affiliation{$^1$Skolkovo Institute of Science and Technology, Bolshoy Boulevard 30, bld.1,
		Moscow, 121205, Russian Federation}
\affiliation{$^2$Department of Applied Mathematics and Theoretical Physics, University of Cambridge, Cambridge CB3 0WA, United Kingdom}

\date{\today}

\begin{abstract}
Strongly coupled light-matter systems can carry information over long distances and realize low threshold
polariton lasing, condensation and superfluidity. These systems are highly non-equilibrium in nature, so constant nonzero fluxes manifest themselves even at the steady state and are set by a complicated interplay between nonlinearity, dispersion, pumping, dissipation and interactions between the various constituents of the system. 
Based on the mean-field governing equations of lasers or polariton condensates, we develop a \nb{method} for engineering  and controlling the velocity profiles by manipulating the spatial pumping and dissipation in the system. We present analytically exact pumping and dissipation profiles that lead to a large variety of spatially periodic density and velocity profiles. \nb{Besides these, any physically relevant  velocity profiles can be engineered by finding the stationary state of the conservative nonlinear Sch\"odinger equation in an external  potential related to the velocity.  Our approach opens the way to the controllable implementation of laser or polariton flows for ultra fast information processing,  integrated circuits, and analogue simulators.}
\end{abstract}

\maketitle

The strong coupling between excitons and photons  results in composite quasi-particles called exciton-polaritons, which have been shown to combine the best properties of their individual components in semiconductor microcavities \cite{Kasprzak_BEC}. Non-interacting and \nb{light} photons greatly reduce the exciton mass by four-five orders of magnitude \cite{deng2003polariton}, whereas excitons contribute their nonlinear properties 
 leading  to the observation of intriguing new physics in solid state systems such as the formation of 
     a macroscopic coherent state 
\cite{Bloch_coherence,Burt_coherence}, polariton condensation \cite{Balili_BEC},  superfluidity \cite{Amo_superfluid1,Amo_superfluid2}, quantized vortices  \cite{Lagoudakis_vortices}, topological effects
\cite{Karzig_topological,Nalitov_topological,lubatsch2019behavior}, and non-equilibrium pattern-forming properties \cite{Wouters_nonequilibrium,Szymanska_nonequilibrium}. Polaritons  are strong candidates for technological applications such as    dissipation-less optical devices \cite{de2012control,marsault2015realization}, analog simulators \cite{berloff2017realizing}, transistors \cite{Gao_transistor,Ballarini_transistor,zasedatelev2019room} and optical switches \cite{Amo_switch,frank2012coherent,frank2013non}.

\begin{figure}[!h]
	\begin{minipage}[c]{\linewidth}
		\includegraphics[width=0.65\linewidth]{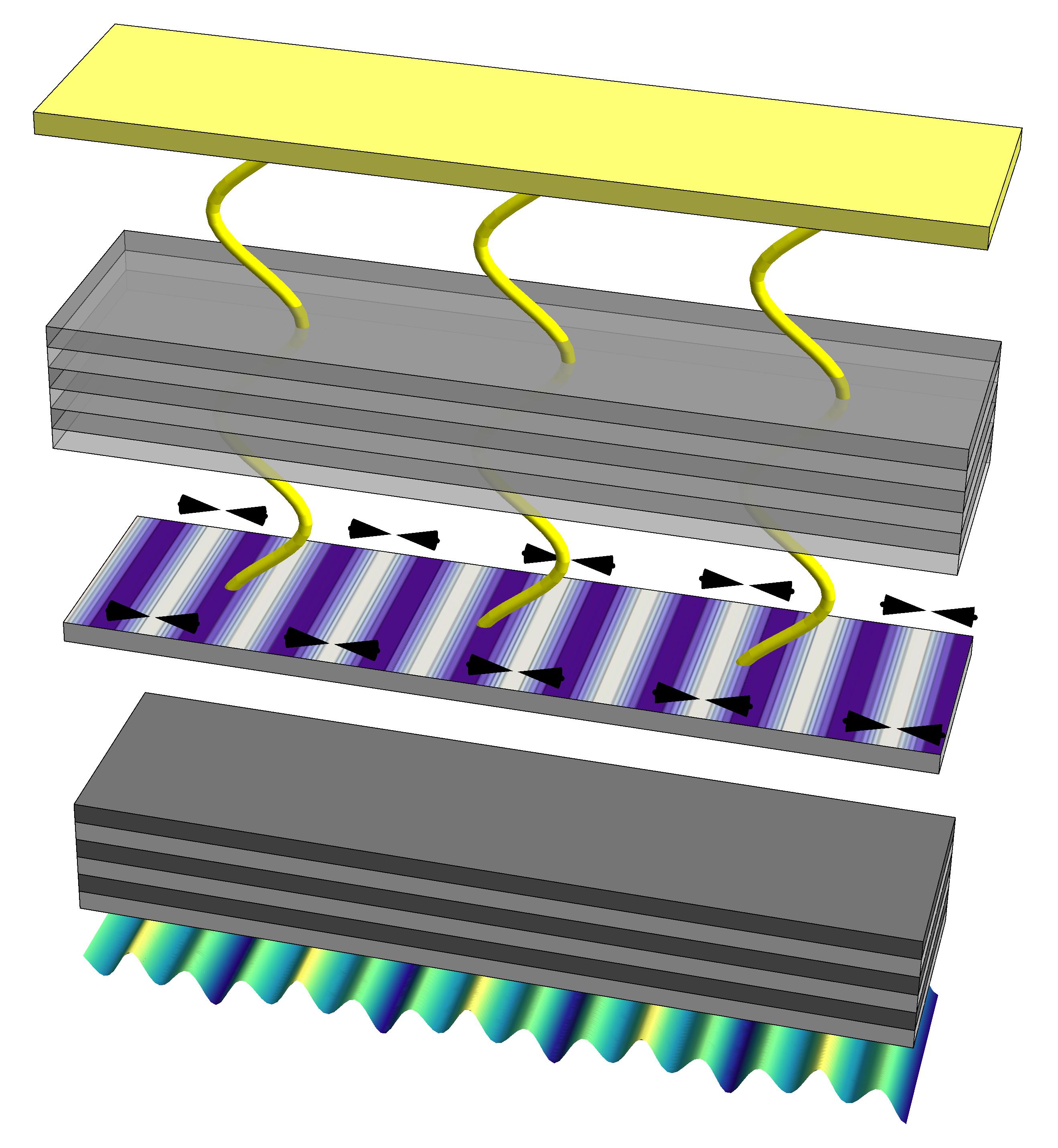}
		\caption{\nb{Schematics of the flow management approach presented in our paper. The top yellow plate represents the source of the incoherent uniform pumping (cw excitation). Grey layers denote the distributed Bragg reflectors between which the quantum well is located and the formation of the polariton condensates takes place. On the top  we show density of the polaritons with the light-blue color scheme  representing the periodicity of  the density regions (dark blue (white) color  marks the maxima (minima) of the density) . The target flow profile is shown by the black arrows  indicating the direction of the flow. Below the structure we show the spatially varying dissipation  that  leads to such a flow profile.}}
		\label{schematic}
	\end{minipage}
	\hfill
\end{figure}

\nb{The precise implementation of well-controllable lattice potentials for bosonic condensates is an inevitable step towards the realization of advanced classical \cite{ozawa2019topological,berloff2017realizing,kalinin2019towards} and quantum simulators \cite{na2010massive}. Polariton condensates in lattices could be manipulated into soliton, which is seen as a building block for polaritonic circuits, where propagation and localization are optically controlled and reconfigurable \cite{tanese2013polariton}. Velocity control is needed to form topologically protected stated (e.g. chiral edge states) at the surface of topological insulators that allow unidirectional transport immune to backscattering
and topological edge modes \cite{st2017lasing},  or to form optical delay lines with enhanced transport properties \cite{mittal2014topologically}.} The full exploitation of  polariton's potential for  technological applications
 requires engineering microstructures with certain high quality trapping profiles  with precise parameters and realising quasiparticles flow and density control. 
 Polariton condensation has been recently achieved in a one-dimensional strong lead halide perovskite lattices   \cite{su2018room,su2019observation} paving the way  for all-optical integrated logic circuits  operating at room temperature. 
  However, unlike optical lattices of equilibrium ultra-cold BECs, the periodic trapping of nonequilibrium condensates provides  flows between the lattice sites that are difficult to predict and challenging to control. In this Letter, we show how to control the velocity and density of polariton condensates by using spatially varying dissipation and pumping profiles. \nb{The spatially controlled dissipation can be achieved by many methods including proton implant technique \cite{Schneider_landscape},  excited states absorption \cite{sanvitto2016road}, by changing the thickness of the film 
\cite{rajendran2019low} and other techniques \cite{kalinin2019towards}. }
Figure \ref{schematic} shows the schematic of the experimental setup with uniform incoherent injection and nonuniform spatially varying dissipation that \nb{creates}  the desired flow pattern.

Our starting point is  a generic laser or nonequilibrium condensate model in the form of the complex Ginzburg-Landau equation (cGLE) with saturable nonlinearity that results from the Maxwell-Bloch equations:
\begin{equation}
i\frac{\partial \Psi}{\partial t} = -\frac{1}{2}\Delta \Psi + | \Psi |^{2} \Psi + gn_{R}\Psi + i[n_{R}-\gamma({\bf x})]\Psi, \label{cGLE1} 
\end{equation}
where $\Psi({\bf x},t)$ is the complex-valued order parameter, $\gamma({\bf x})$ is the spatially varying linear losses due to imperfect confinement, $n_R$ is the density distribution of  noncondensed  reservoir particles (such as hot excitons) that provides  gain saturation $n_R= P({\bf x},t)(1+b|\Psi|^2)^{-1}$, $P({\bf x},t)$ is the gain, $b$ characterizes the relative strengths of the gain saturation and self-interactions. Parameter $g$ characterises the strength of the interactions between the condensate and the reservoir of noncondensed particles.  This model has been quite successful in describing many aspects of lasers and  polariton condensate dynamics  and the resulting steady states \cite{Wouters_nonequilibrium,Szymanska_nonequilibrium, keeling2008spontaneous} due to its universality \cite{Newell_order,Pismen_order,Cross_order}. 

As the pumping is increased above the threshold for condensation, the coherence across the pumping region is established. Close to the condensation threshold and with the constant pumping $P=P_c$, $n_R \approx P_c - P_c b |\Psi|^2$ and Eq.~(\ref{cGLE1}) can be replaced with a more standard form of the cGLE
\begin{equation}
 i\frac{\partial \tilde\Psi}{\partial t} = -\frac{1}{2}\Delta \tilde\Psi +\beta  | \tilde\Psi  |^{2} \tilde\Psi  + i[P_c-\gamma({\bf x})-\sigma|\tilde\Psi|^2]\tilde\Psi,  \label{cGLE}
\end{equation}
where $\sigma=P_cb|1-gP_c b|^{-1}$, $\tilde \Psi=\Psi \exp[i g P_c t]|1-g P_c b|^{1/2},$  and $\beta={\rm sgn} (1-g bP_c)$ characterises the sign of the self-interactions: defocusing ($\beta=+1$) or focusing ($\beta=-1$). We also consider $\beta=0,$ in which case $\sigma=P_cb$ and $\tilde \Psi=\Psi \exp[i g P_c t]$.

If the pumping strength remains constant, the system reaches the steady state characterised by the chemical potential $\mu$, such that $\tilde\Psi=\psi({\bf x}) \exp[-i \mu t]$. 
The steady state written using   the Madelung transformation $\psi = \sqrt{\rho({\bf x},t)} e^{iS({\bf x},t)}$ in terms of the density $\rho$ and phase $S$  satisfies
\begin{eqnarray}
\mu&=&\frac{1}{2}\textbf{u}^{2} - \frac{1}{2}\frac{\nabla^2\sqrt{\rho}}{\sqrt{\rho}} +\beta \rho, \label{mu} \\
\frac{1}{2}\nabla\cdot (\rho \textbf{u})&=&(P-\gamma({\bf x})-\sigma\rho) \rho,  \label{cGLE_alt}
\end{eqnarray}
where $\textbf{u} = \nabla S$ is the condensate outflow velocity profile.

The systems out of equilibrium such as lasers or nonequilibrium condensates  are characterised by the existence of nonzero velocity profiles even at the steady states. The steady-state currents connect regions of net gain with those of net loss which form due to the combination of  density-dependent gain rate and  spatial inhomogeneity of either pumping or dissipation. \nb{The superflow velocities} in turn affect the density profile.
The main question we address in this Letter is how to engineer a given velocity profile using the controls available in our system such as the spatially varying  pumping or dissipation. This control will be achieved by observing that nonlinear Eq.~(\ref{mu}) \nb{can be solved yielding } the solutions in terms of $\rho$  for spatially constant pumping and a given  velocity profiles. Such density when substituted into Eq.~(\ref{cGLE_alt}) allows one to specify  the spatially varying dissipation that leads to the given superflow. 
\nb{This is a generic procedure, that can be made to works for any physically relevant target velocity profile. We will illustrate this approach using the velocity profiles that allow to integrate Eq.~(\ref{mu}) exactly leading to fully analytical pumping and dissipation profiles. We also use  this method for  non-periodic velocity profiles and spatially localised excitations that require numerical integration of Eq.~(\ref{mu}). } 
We consider a one-dimensional velocity profile ${\bf u}=u(x)$, \nb{however, the extension to higher dimensions is straightforward.} Equation~ (\ref{mu}) is the second order nonlinear ordinary differential equation which for a given periodic expression for $u(x)$ relates to the steady state of the Gross-Pitaevskii equation (GPE) that describes trapped quasi-one-dimensional  dilute gas of the Bose - Einstein condensate (BEC). The GPE for such gas reads 
\begin{equation}	
i\hat\psi_{t} = - \frac{1}{2}\hat\psi_{xx} + \beta |\hat\psi|^{2} \hat\psi + V(x)\hat\psi,	
\label{NLSE}		
\end{equation}
where $\hat\psi(x,t)$ is the macroscopic wavefunction of the Bose-Einstein condensate and $V(x)$ is the trapping potential. The steady state of Eq.~(\ref{NLSE}) with $\hat\psi=\sqrt{\hat \rho} \exp[i \hat S-i\hat \mu t]$ and a constant phase $\hat S$ satisfies
$
    \hat \mu = V(x) -  (\sqrt{\hat \rho})_{xx}/2\sqrt{\hat \rho} +\beta \rho, 
$ which reduces to Eq.~(\ref{mu}) if we let $2V(x)=u(x)^2,$ $\hat \mu = \mu$ and $\hat \rho=\rho.$ So the trap used in equilibrium condensates can be thought of as the postulated expressions for the velocity profiles. The  density, therefore,  can be obtained similarly to finding the stationary density of equilibrium condensates in a given external potential either analytically (if such solution is known) or numerically by integrating Eq.~(\ref{NLSE}) in imaginary time while renormalizing the wavefunction by the fixed number of particles. Once we know the density profile, Eq.~(\ref{cGLE_alt}) gives the expression for the spatial dissipation that leads to that density and velocity
\begin{equation}
    \gamma(x)=P_c-\sigma \rho - \frac{1}{2} u_x - \frac{1}{2} \rho_x u/\rho. \label{dissipation}
\end{equation}
For some  external potentials  the exact solution of the GPE could be found.\nb{ An important class of such potentials are Jacobi elliptic functions, so we first consider}
$u(x)^2 = - 2V_{0} {\rm sn}^{2} (x,k)$, 
where ${\rm sn}(x,k)$ denotes the Jacobi elliptic sine function with an elliptic parameter $0<k<1$ and $V_0<0$ characterizes the depth of the periodic variation. \nb{ Such  periodic profiles  require translational invariance, that can be realised experimentally using toroidal geometry of excitation \cite{cristofolini2013optical}. }
Denoting $r(x)=\sqrt{\rho(x)}$, Eq.~(\ref{mu}) becomes
\begin{equation}	
\mu r^4 (x) = - \frac{r^3(x)r''(x)}{2} +\beta r^6(x) - V_{0}{\rm sn}^{2}(x,k)r^{4}(x).
\label{density_row}
\end{equation}

 The solutions to the nonlinear ordinary differential Eq.~(\ref{density_row}) can be obtained by expanding $r(x)$ in terms of the various Jacobi elliptic functions (${\rm sn, dn, cn}$), substituting $r(x)$ into Eq.~(\ref{density_row}) and  equating equal powers of these functions to zero to get the expressions for the parameters  \cite{Bronski_attractive,Bronski_repulsive}.

{\it Defocusing regime; $\beta=1$.} Firstly, we consider 
$r^2 (x) = A\, {\rm sn}^2 (x,k) + B,$ which leads to the following  conditions
$\mu = \frac{1}{2}(1+k^2 +3B - \frac{BV_{0}}{k^2-V_{0}}),$ $0 = B(1+\frac{B}{k^2+V_{0}})(k^2+V_{0} + Bk^2),$
$A = k^2 +V_{0}.$ These conditions are satisfied in the following three cases. 1) $B=0$ and $A = k^2+V_{0}$ with $\mu = (1+k^2)/2$ result in
\begin{equation}
	\rho(x) = (V_{0}+k^2) {\rm sn}^2(x,k), \quad -k^2\le V_0 < 0;
\label{rep_A1}
\end{equation}
2) $B=-A = -(k^2+V_{0})$ with $\mu = 1/2 - k^2 - V_{0}$ results in
	$\rho(x) = -(V_{0}+k^2) {\rm cn}^2(x,k),  \quad V_0\le -k^2; $
3)  $B=-(V_{0}+k^2)/k^2 = -A/k^2$ with $\mu = -1 - V_{0}/k^2 + k^2/2$ results in
\begin{equation}
	\rho(x) = \frac{-(V_{0}+k^2)}{k^2} {\rm dn}^2(x,k),  \quad V_0\le -k^2.
\label{rep_A3}
\end{equation}
	
\begin{figure} 
\centering
\includegraphics[width=3.3 in]{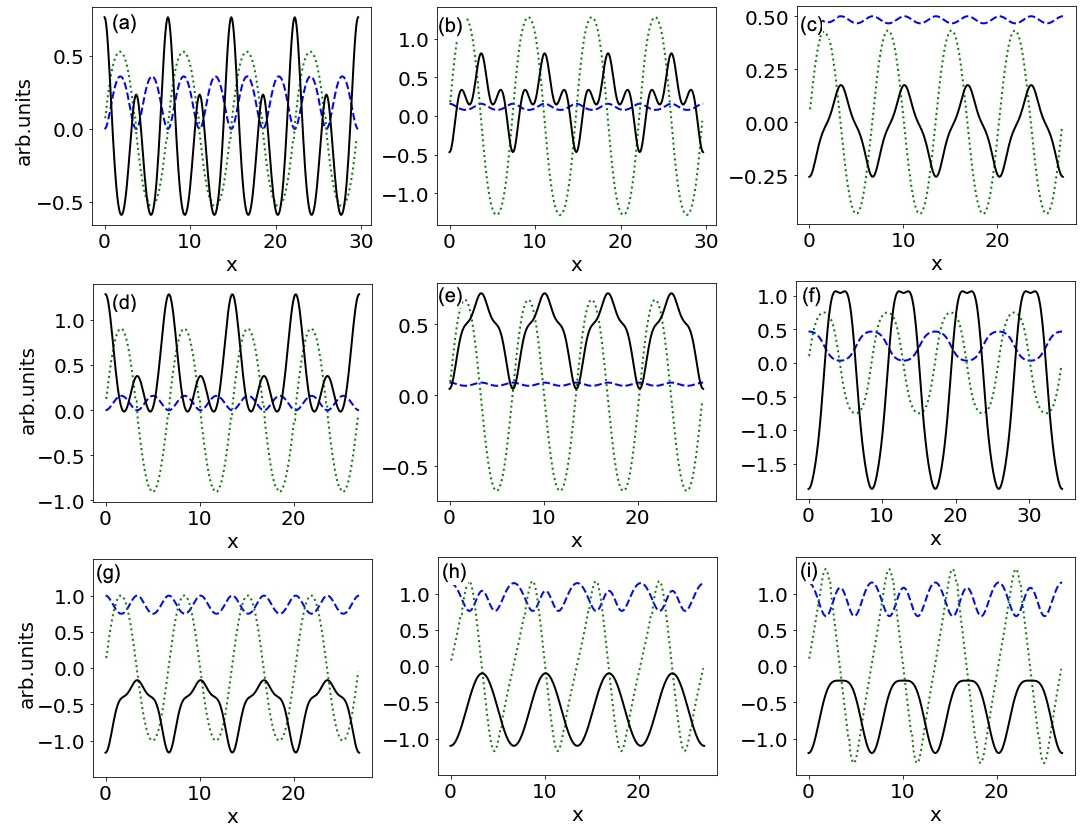}
\caption{Flow velocity $u(x)$ (green dotted lines), density $\rho(x)$ (dashed blue lines) and the dissipation profile $\gamma(x)$ (black solid lines) in the repulsive  $\beta=1$ (a,b,c),(g,h,i) and attractive  $\beta=-1$ (d,e,f) regimes (with $g=1,b=1.5$). Shown are 
			the solutions given by Eq.~(\ref{rep_A1}) with $P_{c} = 0.5,$  $k=0.707,$ $V_{0}=-0.14$ (a), Eq.~(\ref{rep_A3}) with $P_{c} = 1/3,$  $k=0.707,$ $V_{0}=-0.58$ (b),
		    Eq.~(\ref{rep_B23}) with $P_{c} = 0.5,$  $k=0.866,$ $V_{0}=-0.28125$ (c), Eq.~(\ref{att_A1}) with $P_{c} = 5/6,$  $k=0.5,$ $V_{0}=-0.41$ (d), Eq.~(\ref{att_A3}) with $P_{c} = 5/6,$ $k=0.5,$ $V_{0}=-0.2275$ (e),
			Eq.~(\ref{att_B12}) with $P_{c} = 5/6,$  $k=0.866,$ $V_{0}=-0.28125$ (f),
			Eq.~(\ref{rep_A3}) with $P_{c} = 1/3,$  $k=0.5,$ $V_{0}=-0.5$ \nn{(g)}.
            The last row also demonstrates \nn{what happens if the control leading to solution shown in  (g) is perturbed  with  the perturbed dissipation profiles    given by the equations $\gamma=0.5\gamma_{max}(1-{\rm sign}({\rm cn}(x,k))|{\rm cn}(x,k))|^{0.8}$ (h) and $\gamma=0.5\gamma_{max}(1-{\rm sign}({\rm cn}(x,k))|{\rm cn}(x,k))|^{0.6}$ (i).}}
\label{6profiles}
\end{figure}

Secondly, we consider 
$r^2 (x) = a_{1} {\rm cn} (x,k) + b_{1}$ which  yields the following conditions on the coefficients:
$V_{0} = -\frac{3}{8} k^2,$
$\mu = \frac{1}{8}(1+k^2) + \frac{6a_{1}^2}{k^2},$
$0 = \frac{a_{1}^2}{4 k^6}(16a_{1}^2-k^4)(16a_{1}^2+k^2-k^4),$
$b_{1} = \frac{4a_{1}^2}{k^2}$. This leads to two  possibilities for amplitude $a_1$:  $a_{1} = k^2/4,$ and $a_{1} = -k^2/4,$   resulting in
	$\rho(x) = \frac{k^2}{4} (1\pm{\rm cn}(x,k)).$
Finally we consider
$r^2 (x) = a_{2} {\rm dn} (x,k) + b_{2}$ that gives the conditions on the parameters
$V_{0} = -\frac{3}{8} k^2,$
$\mu = \frac{1}{8}(1+k^2) + 6a_{2}^2,$
$0 = \frac{a_{2}^2}{4}(16a_{2}^2-1)(16a_{2}^2+k^2-1),$
$b_{2} = 4a_{2}^2$. 
This leads to 
 $a_{2} = \pm1/4$, $\mu = 1/2 + k^2/8$ with the expression on the amplitude
	$\rho(x) = \frac{1}{4}(1 \pm {\rm dn}(x,k)),$
or
2) $a_{2} = \sqrt{1-k^2}/4,$ $\mu = 1-k^2/2$ with $k'=\sqrt{1-k^2}$ leading to
\begin{equation}
	\rho(x) = \frac{k'}{4}(k' + {\rm dn}(x,k)).
\label{rep_B23}
\end{equation}
The dissipation profile that leads to these density and flow profiles is determined by Eq.~(\ref{dissipation}). The representative examples of these solutions are depicted in Fig.~\ref{6profiles}\nn{(a,b,c)}.
	
{\it Focusing regime; $\beta=-1$.}  The focusing case yields some different sets of solutions. Taking $r^2 (x) = A {\rm sn}^2 (x,k) + B$ leads to
$\mu = \frac{1}{2}(1+k^2 -3B + \frac{BV_{0}}{V_{0}+k^2}),$
$0 = B(\frac{B}{V_{0} + k^2}-1)(V_{0} + k^2 - Bk^2),$
$A = -(V_{0} + k^2).$ The  possible choices are
1) $B=0$ with $\mu = (1+k^2)/2$ leading to
\begin{equation}
	\rho(x) = -(V_{0}+k^2) {\rm sn}^2(x,k),\quad -k^2\ge V_0,
\label{att_A1}
\end{equation}
2) $B=- (k^2+V_{0})$,  with $\mu = 1/2 - k^2 - V_{0}$ leading to
	$\rho(x) = (V_{0}+k^2) {\rm cn}^2(x,k),\quad -k^2\le V_0< 0,$ 
or
3) $B = (V_{0}+k^2)/k^2$ with $\mu = -1 - V_{0}/k^2 + k^2/2$ leading to 
\begin{equation}
	\rho(x) = \frac{(V_{0}+k^2)}{k^2} {\rm dn}^2(x,k), \quad -k^2\le V_0< 0.
\label{att_A3}
\end{equation}
Taking
$r^2 (x) = a_{1} {\rm sn} (x,k) + b_{1}$  gives us a set of conditions:
$V_{0} = -\frac{3}{8} k^2,$
$\mu = \frac{1}{8}(1+k^2) - \frac{6a_{1}^2}{k^2},$
$0 = -\frac{a_{1}^2}{4 k^6}(16a_{1}^2-k^4)(16a_{1}^2-k^2),$
$b_{1} = \frac{4a_{1}^2}{k^2}.$ If 
1) $a_{1} = \pm k^2/4$, $\mu = 1/8 - k^2/4,$  we get
	$\rho(x) = \frac{k^2}{4}(1 \pm {\rm sn}(x,k)),$
or   2) $a_{1} = \pm k/4$, $\mu = -1/4 + k^2/8,$  so that 
\begin{equation}
	\rho(x) = \frac{1}{4}(1 \pm k\, {\rm sn}(x,k)). 
\label{att_B12}
\end{equation}
Finally, we consider
$r^2 (x) = a_{2} {\rm dn} (x,k) + b_{2}$ which yields
$V_{0} = -\frac{3}{8} k^2,$
$\mu = \frac{1}{8}(1+k^2) + 6a_{2}^2,$
$0 = \frac{a_{2}^2}{4}(16a_{2}^2-1)(16a_{2}^2+k^2-1),$ and
$b_{2} = -4a_{2}^2$.
The only nontrivial solution exists for $a_{2} = \sqrt{1-k^2}/4.$ Let $k'=\sqrt{1-k^2},$ and with $\mu = 1/4 + k'^2/4$  the solution becomes
	$\rho(x) = \frac{k'}{4}({\rm dn}(x,k) - k').$
For any given uniform pumping $P_c$ the corresponding dissipative profile is given by Eq.~(\ref{dissipation}) with some representative examples of these solutions  shown in Fig.~\ref{6profiles}\nn{(d,e,f)}.

{\it Non-interacting case: $\beta=0.$}
The case of $g P_c b =1$  has fewer physically relevant solutions.
For $r^2 (x) = A {\rm sn}^2 (x,k) + B$, we get
$2\mu AB = -V_{0} B^2 + AB + ABk^2,$
$0 = (\mu B + A/2)B,$
$2 \mu A^2 = -3 A k^2 B + A^2 (1+k^2) -4 ABV_{0},$
$V_{0} = -k^2,$ which leads to the following possibilities.
1)$B=0$ and $\mu = (1+k^2)/2$ gives (with $A>0$):
	$\rho(x) = A\, {\rm sn}^2(x,k),$
or 2) $2\mu = -A/B = 1$ which yields ($A<0$):
	$\rho(x) = - A \,{\rm cn}^2(x,k),$
and 
3) $2\mu = -A/B = k^2$  results in ($A<0$):
\begin{equation}
	\rho(x) = -A \,{\rm dn}^2(x,k)/k^2.
\label{zero_A3}
\end{equation}
For $r^2 (x) = a_{1} {\rm cn} (x,k) + b_{1}$, or $r^2 (x) = a_{1} {\rm dn} (x,k) + b_{1}$  setting the terms before the basic functions to zero imposes equations:
$\mu a_1^2 = -V_{0} a_1^2 + V_{0} b_1^2 + (1 - 2 k^2)a_1^2/8,$
$\mu b_1^2 = -V_{0} b_1^2 + (1 - k^2)a_1^2/8,$ for {\rm cn}-type and
$\mu a_1^2 = -V_{0} a_1^2/k^2 + V_{0} b_1^2/k^2 + (k^2 - 2)a_1^2/8,$
$\mu b_1^2 = -V_{0} b_1^2/k^2 + (k^2 - 1)a_1^2/8,$ for {\rm dn}-type with two common equations
$V_{0} a_1^2 + (3 a_1^2 k^2)/8  = 0,$ and
$4 V_{0} a_1b_1 + a_1b_1k^2 = 0,$ which are compatable only with $b_{1}=0,$ $V_{0}=-3k^2/8,$ giving the resulting values for
$k=1,$ $\mu=1/4$. The final solutions in both cases take the form of a solitary wave
	$\rho(x) = a_{1} {\rm sech(x)}, a_1>0.$

\begin{figure}
\centering

\includegraphics[width=3.3in]{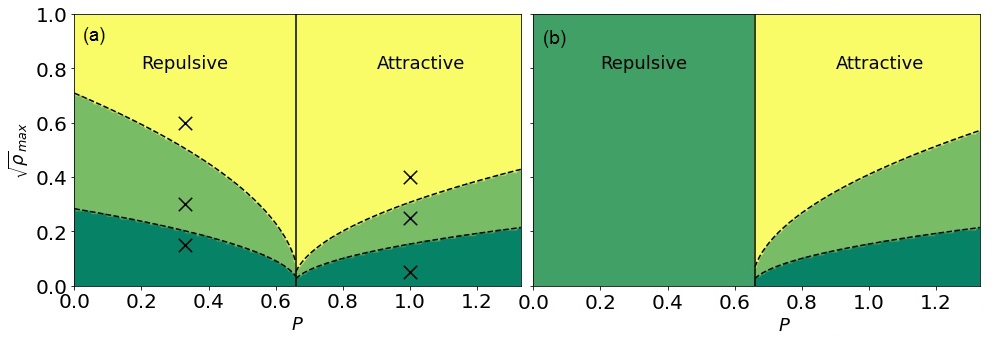} 
\caption{Stability diagrams for the family of {\rm dn}\textit{(x,k)} solutions given by Eq.~(\ref{rep_A3}), Eq.~(\ref{att_A3}) and Eq.~(\ref{zero_A3}) (with $g=1,b=1.5$).obtained by numerical integration of Eq.~(\ref{cGLE1})(a) and Eq.~(\ref{cGLE})(b). Green color stands for stable, yellow - for unstable and {light green} - for the transitional regimes to a different stable configuration. Black line depicts the solutions with $\beta=0$, {dash lines - the borders for the specified regions, while black crosses indicate the solutions also shown in Fig.~\ref{3evolution}.}}
\label{phase_diagram}
\end{figure}

Next, we consider the stability of the presented solutions by evolving them dynamically according to Eq.~(\ref{cGLE1}) or the full Eq.~(\ref{cGLE}). \nn{Figure \ref{phase_diagram} summarises our findings using the family of {\rm dn}\textit{(x,k)} solutions given by Eq.~(\ref{rep_A3}), Eq.~(\ref{att_A3}) and Eq.~(\ref{zero_A3}).} The instability takes place when the dissipative profile becomes strongly negative in some spatial regions, so starts representing the generation of quasiparticles that can not be compensated by the remaining positive dissipative regions. This leads to the uncontrollable growth of particles in the repulsive ($\beta=1$) and non-interactive ($\beta=0$) regimes; in attractive  regime ($\beta=-1$) the system either exhibits oscillations  due to the interplay between attractive forces and fast saturation response or undergoes the transition to another stationary solution. \nb{We note that when the solution is stable, the system finds it starting from any random noise, not just from the perturbed solution. This assures that the control is working and leads to the target velocity profile.}

\begin{figure}
\centering
\includegraphics[width=3.3 in]{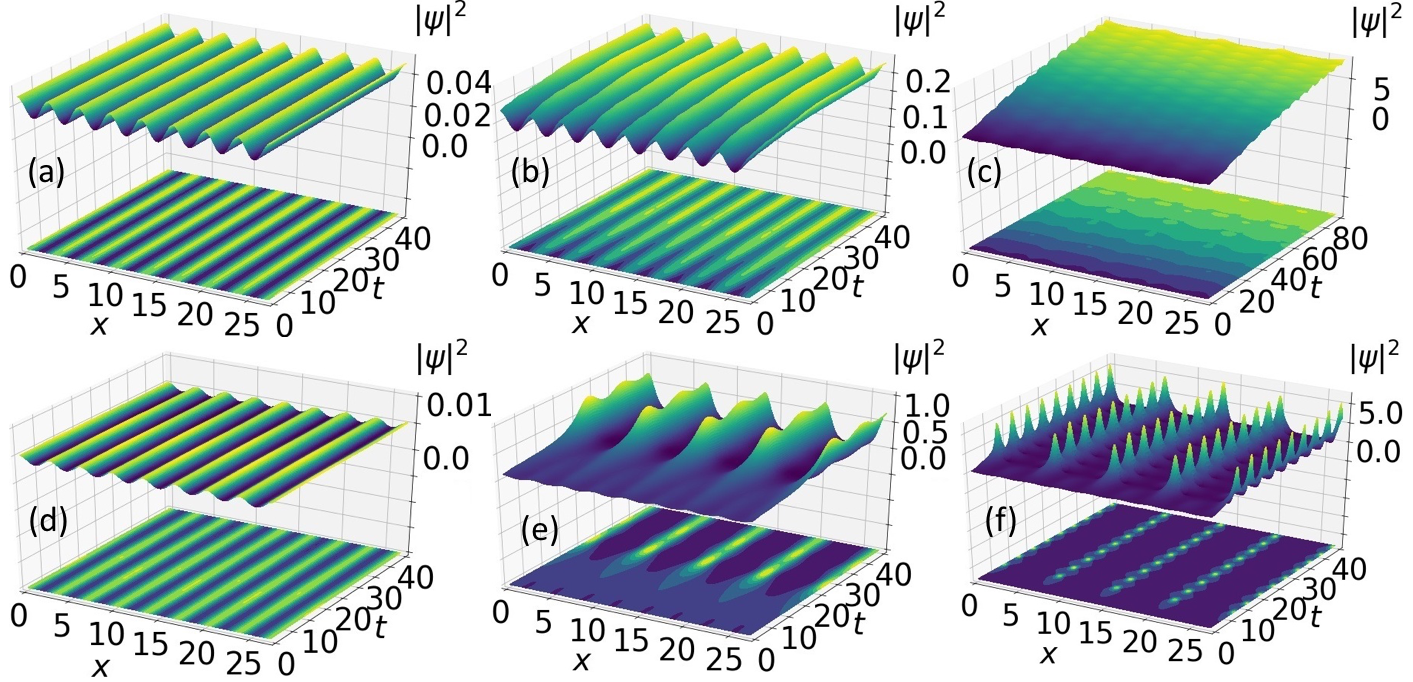} 
\caption{The stable (left column), transitional (middle column), unstable (right column) evolutions of  {\rm dn}\textit{(x,k)} solutions given by Eq.~(\ref{rep_A3}) in a repulsive regime ($P_{c} = 1/3,$ $\sigma = 1, \beta=1, g=1$)  with $k=0.5$ and  $\sqrt{\rho_{max}}(t=0)=0.15$ (a), $\sqrt{\rho_{max}}(t=0)=0.3$ (b), $\sqrt{\rho_{max}}(t=0)=0.6$ (c) and by   {\rm dn}\textit{(x,k)} solutions given by Eq.~(\ref{att_A3}) in an attractive regime ($P_{c} = 1,$ $\sigma = 3, \beta=-1$) with $k=0.5$ and 
 and  $\sqrt{\rho_{\rm max}}(t=0)=0.05$ (d), $\sqrt{\rho_{\rm max}}(t=0)=0.25$ (e),  $\sqrt{\rho_{\rm max}}(t=0)=0.4$ (f).}
\label{3evolution}
\end{figure}

 Figures \ref{3evolution} (a-c) and (d-f) demonstrate the  behavior of the solutions governed by Eq.~(\ref{rep_A3}) and Eq.~(\ref{att_A3}) respectively: \nn{stable  evolution (Fig.\ref{3evolution}(a,d)), evolution to a different stationary solution (Fig.\ref{3evolution}(b,e)) and either unlimited growth of the density due to the lack of the gain saturation (Fig.\ref{3evolution}(c)) or time-periodic oscillations (Fig.\ref{3evolution}(f))}. 
 
 
On the one hand, the negative dissipation (gain) has been  experimentally realised  by using injection below the threshold for the condensate formation \cite{alyatkin2020optical}. On the other hand, one can use the linear relationship between the pumping intensity $P_c$ and the dissipation given by Eq. (\ref{dissipation}) to increase $P_c$ to make $\gamma>0$ for all ${\bf x}$. This will stabilise the solution while removing the necessity to generate nonuniform gain. This will not work for analytic solution if changing  $P_c$ changes the sign of  $\beta = {\rm sign} (1-P_c g b)$ that controls the type of the solution (focusing, defocusing, linear). So this approach would not work if increase in $P_c$ changes the solution type. To overcome this problem one may choose the microcavity with lower values of $g$ (more excitonic) and $b$ (stronger polariton-polariton interactions)\cite{estrecho2019direct}. In Figs. 2-4 we used the experimentally largest values of  $g$ and $b$.

\nb{Experimentally achievable dissipation and pump may  suffer from imperfections. In Fig.\ref{6profiles} (g,h,i) we show the effect of perturbing  the effective pumping/dissipation profiles.}

\begin{figure} 
\centering
\includegraphics[width=2.5in]{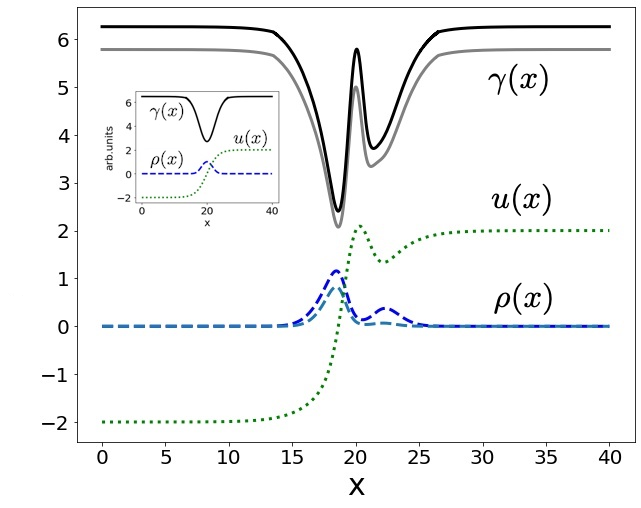} 
\caption{\nb{Flow velocity $u(x)$ (green dotted lines), density $\rho(x)$ (dashed dark and white blue lines) and the dissipation profile $\gamma(x)$ (black and grey solid lines) in the repulsive, $\beta=1,$ regime. 
The velocity profile and pumping were   chosen as $u(x)=2\tanh[0.3(x-20)]+2\exp[-0.5(x-20)^2]$   and $P_c=4$. The number of the particles as  $N = \int |\Psi|^2 dx = 2$  (faint dashed-blue line for density and  black line for dissipation) and $N = 4$  (dark-blue line for density and grey line for dissipation). The inset shows the velocity profile that leads to a single localised condensate peak and the velocity profile $u(x)=2\tanh(0.3(x-20))$ .}}
\label{profiles_with_inset}
\end{figure}

\nb{{\it General framework for the velocity engineering. } So far we considered the velocity profiles for which the analytical solution of the governing equations exists: periodic profiles are expressed as  Jacobi elliptic functions. However, { \it any}  physically realistic velocity profile can be engineered using numerical integration of Eq.~(\ref{NLSE}).  Some of the resulting pumping/dissipation profiles are shown in Fig.~\ref{profiles_with_inset} for analytical velocity profiles that correspond to a spatially  localised excitations. Several subtleties arise that are not present when analytical solutions of Eq.~(\ref{NLSE}) exist. To find the steady state of Eq.~(\ref{mu}) for a given velocity profile, we used the relaxation of   Eq.~(\ref{NLSE}) (integration in imaginary time) while renormalizing the wavefunction to a fixed number of particles $N$ at every time step. To use the spatially localised excitation, the velocity profile has to be a constant away from the excitation spot \cite{lagoudakis2017polariton}, so we used the velocity profiles that satisfy this condition. The resulting density decays as $\exp[-2 \mu x]$ away from the excitation to satisfy Eq.~(\ref{mu}). Such straightforward asymptotics allows us to evaluate the last term of Eq.~(\ref{dissipation}) that otherwise would be problematic  to find numerically  for small densities.
}

The cGLE-type systems are pattern-forming systems with a complex interplay of gain, dissipation, nonlinearities due to  self interactions and gain saturation. This makes the engineering of the flow and density profiles in such systems challenging. However, such control is necessary for implementation of various proposals from topological insulators to optical/photonic/polaritonic transistors, information processing devices \nb{and Hamiltonian simulators. In this Letter we proposed a method for realising periodic modulations of the density and velocity profiles in a wide variety of laser/condensate systems governed by the cGLE.}

The authors  acknowledge funding from Huawei project 1-10-3292.

\bibliography{Bibliography}{}

\begin{thebibliography}{43}%
\makeatletter
\providecommand \@ifxundefined [1]{%
 \@ifx{#1\undefined}
}%
\providecommand \@ifnum [1]{%
 \ifnum #1\expandafter \@firstoftwo
 \else \expandafter \@secondoftwo
 \fi
}%
\providecommand \@ifx [1]{%
 \ifx #1\expandafter \@firstoftwo
 \else \expandafter \@secondoftwo
 \fi
}%
\providecommand \natexlab [1]{#1}%
\providecommand \enquote  [1]{``#1''}%
\providecommand \bibnamefont  [1]{#1}%
\providecommand \bibfnamefont [1]{#1}%
\providecommand \citenamefont [1]{#1}%
\providecommand \href@noop [0]{\@secondoftwo}%
\providecommand \href [0]{\begingroup \@sanitize@url \@href}%
\providecommand \@href[1]{\@@startlink{#1}\@@href}%
\providecommand \@@href[1]{\endgroup#1\@@endlink}%
\providecommand \@sanitize@url [0]{\catcode `\\12\catcode `\$12\catcode
  `\&12\catcode `\#12\catcode `\^12\catcode `\_12\catcode `\%12\relax}%
\providecommand \@@startlink[1]{}%
\providecommand \@@endlink[0]{}%
\providecommand \url  [0]{\begingroup\@sanitize@url \@url }%
\providecommand \@url [1]{\endgroup\@href {#1}{\urlprefix }}%
\providecommand \urlprefix  [0]{URL }%
\providecommand \Eprint [0]{\href }%
\providecommand \doibase [0]{http://dx.doi.org/}%
\providecommand \selectlanguage [0]{\@gobble}%
\providecommand \bibinfo  [0]{\@secondoftwo}%
\providecommand \bibfield  [0]{\@secondoftwo}%
\providecommand \translation [1]{[#1]}%
\providecommand \BibitemOpen [0]{}%
\providecommand \bibitemStop [0]{}%
\providecommand \bibitemNoStop [0]{.\EOS\space}%
\providecommand \EOS [0]{\spacefactor3000\relax}%
\providecommand \BibitemShut  [1]{\csname bibitem#1\endcsname}%
\let\auto@bib@innerbib\@empty
\bibitem [{\citenamefont {Kasprzak}\ \emph {et~al.}(2006)\citenamefont
  {Kasprzak}, \citenamefont {Richard}, \citenamefont {Kundermann},
  \citenamefont {Baas}, \citenamefont {Jeambrun}, \citenamefont {Keeling},
  \citenamefont {Marchetti}, \citenamefont {Szyma{\'n}ska}, \citenamefont
  {Andr{\'e}}, \citenamefont {Staehli} \emph {et~al.}}]{Kasprzak_BEC}%
  \BibitemOpen
  \bibfield  {author} {\bibinfo {author} {\bibfnamefont {J.}~\bibnamefont
  {Kasprzak}}, \bibinfo {author} {\bibfnamefont {M.}~\bibnamefont {Richard}},
  \bibinfo {author} {\bibfnamefont {S.}~\bibnamefont {Kundermann}}, \bibinfo
  {author} {\bibfnamefont {A.}~\bibnamefont {Baas}}, \bibinfo {author}
  {\bibfnamefont {P.}~\bibnamefont {Jeambrun}}, \bibinfo {author}
  {\bibfnamefont {J.}~\bibnamefont {Keeling}}, \bibinfo {author} {\bibfnamefont
  {F.}~\bibnamefont {Marchetti}}, \bibinfo {author} {\bibfnamefont
  {M.}~\bibnamefont {Szyma{\'n}ska}}, \bibinfo {author} {\bibfnamefont
  {R.}~\bibnamefont {Andr{\'e}}}, \bibinfo {author} {\bibfnamefont
  {J.}~\bibnamefont {Staehli}},  \emph {et~al.},\ }\href@noop {} {\bibfield
  {journal} {\bibinfo  {journal} {Nature}\ }\textbf {\bibinfo {volume} {443}},\
  \bibinfo {pages} {409} (\bibinfo {year} {2006})}\BibitemShut {NoStop}%
\bibitem [{\citenamefont {Deng}\ \emph {et~al.}(2003)\citenamefont {Deng},
  \citenamefont {Weihs}, \citenamefont {Snoke}, \citenamefont {Bloch},\ and\
  \citenamefont {Yamamoto}}]{deng2003polariton}%
  \BibitemOpen
  \bibfield  {author} {\bibinfo {author} {\bibfnamefont {H.}~\bibnamefont
  {Deng}}, \bibinfo {author} {\bibfnamefont {G.}~\bibnamefont {Weihs}},
  \bibinfo {author} {\bibfnamefont {D.}~\bibnamefont {Snoke}}, \bibinfo
  {author} {\bibfnamefont {J.}~\bibnamefont {Bloch}}, \ and\ \bibinfo {author}
  {\bibfnamefont {Y.}~\bibnamefont {Yamamoto}},\ }\href@noop {} {\bibfield
  {journal} {\bibinfo  {journal} {Proceedings of the National Academy of
  Sciences}\ }\textbf {\bibinfo {volume} {100}},\ \bibinfo {pages} {15318}
  (\bibinfo {year} {2003})}\BibitemShut {NoStop}%
\bibitem [{\citenamefont {Bloch}\ \emph {et~al.}(2000)\citenamefont {Bloch},
  \citenamefont {H{\"a}nsch},\ and\ \citenamefont
  {Esslinger}}]{Bloch_coherence}%
  \BibitemOpen
  \bibfield  {author} {\bibinfo {author} {\bibfnamefont {I.}~\bibnamefont
  {Bloch}}, \bibinfo {author} {\bibfnamefont {T.~W.}\ \bibnamefont
  {H{\"a}nsch}}, \ and\ \bibinfo {author} {\bibfnamefont {T.}~\bibnamefont
  {Esslinger}},\ }\href@noop {} {\bibfield  {journal} {\bibinfo  {journal}
  {Nature}\ }\textbf {\bibinfo {volume} {403}},\ \bibinfo {pages} {166}
  (\bibinfo {year} {2000})}\BibitemShut {NoStop}%
\bibitem [{\citenamefont {Burt}\ \emph {et~al.}(1997)\citenamefont {Burt},
  \citenamefont {Ghrist}, \citenamefont {Myatt}, \citenamefont {Holland},
  \citenamefont {Cornell},\ and\ \citenamefont {Wieman}}]{Burt_coherence}%
  \BibitemOpen
  \bibfield  {author} {\bibinfo {author} {\bibfnamefont {E.}~\bibnamefont
  {Burt}}, \bibinfo {author} {\bibfnamefont {R.}~\bibnamefont {Ghrist}},
  \bibinfo {author} {\bibfnamefont {C.}~\bibnamefont {Myatt}}, \bibinfo
  {author} {\bibfnamefont {M.}~\bibnamefont {Holland}}, \bibinfo {author}
  {\bibfnamefont {E.~A.}\ \bibnamefont {Cornell}}, \ and\ \bibinfo {author}
  {\bibfnamefont {C.}~\bibnamefont {Wieman}},\ }\href@noop {} {\bibfield
  {journal} {\bibinfo  {journal} {Physical Review Letters}\ }\textbf {\bibinfo
  {volume} {79}},\ \bibinfo {pages} {337} (\bibinfo {year} {1997})}\BibitemShut
  {NoStop}%
\bibitem [{\citenamefont {Balili}\ \emph {et~al.}(2007)\citenamefont {Balili},
  \citenamefont {Hartwell}, \citenamefont {Snoke}, \citenamefont {Pfeiffer},\
  and\ \citenamefont {West}}]{Balili_BEC}%
  \BibitemOpen
  \bibfield  {author} {\bibinfo {author} {\bibfnamefont {R.}~\bibnamefont
  {Balili}}, \bibinfo {author} {\bibfnamefont {V.}~\bibnamefont {Hartwell}},
  \bibinfo {author} {\bibfnamefont {D.}~\bibnamefont {Snoke}}, \bibinfo
  {author} {\bibfnamefont {L.}~\bibnamefont {Pfeiffer}}, \ and\ \bibinfo
  {author} {\bibfnamefont {K.}~\bibnamefont {West}},\ }\href@noop {} {\bibfield
   {journal} {\bibinfo  {journal} {Science}\ }\textbf {\bibinfo {volume}
  {316}},\ \bibinfo {pages} {1007} (\bibinfo {year} {2007})}\BibitemShut
  {NoStop}%
\bibitem [{\citenamefont {Amo}\ \emph {et~al.}(2009{\natexlab{a}})\citenamefont
  {Amo}, \citenamefont {Lefr{\`e}re}, \citenamefont {Pigeon}, \citenamefont
  {Adrados}, \citenamefont {Ciuti}, \citenamefont {Carusotto}, \citenamefont
  {Houdr{\'e}}, \citenamefont {Giacobino},\ and\ \citenamefont
  {Bramati}}]{Amo_superfluid1}%
  \BibitemOpen
  \bibfield  {author} {\bibinfo {author} {\bibfnamefont {A.}~\bibnamefont
  {Amo}}, \bibinfo {author} {\bibfnamefont {J.}~\bibnamefont {Lefr{\`e}re}},
  \bibinfo {author} {\bibfnamefont {S.}~\bibnamefont {Pigeon}}, \bibinfo
  {author} {\bibfnamefont {C.}~\bibnamefont {Adrados}}, \bibinfo {author}
  {\bibfnamefont {C.}~\bibnamefont {Ciuti}}, \bibinfo {author} {\bibfnamefont
  {I.}~\bibnamefont {Carusotto}}, \bibinfo {author} {\bibfnamefont
  {R.}~\bibnamefont {Houdr{\'e}}}, \bibinfo {author} {\bibfnamefont
  {E.}~\bibnamefont {Giacobino}}, \ and\ \bibinfo {author} {\bibfnamefont
  {A.}~\bibnamefont {Bramati}},\ }\href@noop {} {\bibfield  {journal} {\bibinfo
   {journal} {Nature Physics}\ }\textbf {\bibinfo {volume} {5}},\ \bibinfo
  {pages} {805} (\bibinfo {year} {2009}{\natexlab{a}})}\BibitemShut {NoStop}%
\bibitem [{\citenamefont {Amo}\ \emph {et~al.}(2009{\natexlab{b}})\citenamefont
  {Amo}, \citenamefont {Sanvitto}, \citenamefont {Laussy}, \citenamefont
  {Ballarini}, \citenamefont {Del~Valle}, \citenamefont {Martin}, \citenamefont
  {Lemaitre}, \citenamefont {Bloch}, \citenamefont {Krizhanovskii},
  \citenamefont {Skolnick} \emph {et~al.}}]{Amo_superfluid2}%
  \BibitemOpen
  \bibfield  {author} {\bibinfo {author} {\bibfnamefont {A.}~\bibnamefont
  {Amo}}, \bibinfo {author} {\bibfnamefont {D.}~\bibnamefont {Sanvitto}},
  \bibinfo {author} {\bibfnamefont {F.}~\bibnamefont {Laussy}}, \bibinfo
  {author} {\bibfnamefont {D.}~\bibnamefont {Ballarini}}, \bibinfo {author}
  {\bibfnamefont {E.}~\bibnamefont {Del~Valle}}, \bibinfo {author}
  {\bibfnamefont {M.}~\bibnamefont {Martin}}, \bibinfo {author} {\bibfnamefont
  {A.}~\bibnamefont {Lemaitre}}, \bibinfo {author} {\bibfnamefont
  {J.}~\bibnamefont {Bloch}}, \bibinfo {author} {\bibfnamefont
  {D.}~\bibnamefont {Krizhanovskii}}, \bibinfo {author} {\bibfnamefont
  {M.}~\bibnamefont {Skolnick}},  \emph {et~al.},\ }\href@noop {} {\bibfield
  {journal} {\bibinfo  {journal} {Nature}\ }\textbf {\bibinfo {volume} {457}},\
  \bibinfo {pages} {291} (\bibinfo {year} {2009}{\natexlab{b}})}\BibitemShut
  {NoStop}%
\bibitem [{\citenamefont {Lagoudakis}\ \emph {et~al.}(2008)\citenamefont
  {Lagoudakis}, \citenamefont {Wouters}, \citenamefont {Richard}, \citenamefont
  {Baas}, \citenamefont {Carusotto}, \citenamefont {Andr{\'e}}, \citenamefont
  {Dang},\ and\ \citenamefont {Deveaud-Pl{\'e}dran}}]{Lagoudakis_vortices}%
  \BibitemOpen
  \bibfield  {author} {\bibinfo {author} {\bibfnamefont {K.~G.}\ \bibnamefont
  {Lagoudakis}}, \bibinfo {author} {\bibfnamefont {M.}~\bibnamefont {Wouters}},
  \bibinfo {author} {\bibfnamefont {M.}~\bibnamefont {Richard}}, \bibinfo
  {author} {\bibfnamefont {A.}~\bibnamefont {Baas}}, \bibinfo {author}
  {\bibfnamefont {I.}~\bibnamefont {Carusotto}}, \bibinfo {author}
  {\bibfnamefont {R.}~\bibnamefont {Andr{\'e}}}, \bibinfo {author}
  {\bibfnamefont {L.~S.}\ \bibnamefont {Dang}}, \ and\ \bibinfo {author}
  {\bibfnamefont {B.}~\bibnamefont {Deveaud-Pl{\'e}dran}},\ }\href@noop {}
  {\bibfield  {journal} {\bibinfo  {journal} {Nature Physics}\ }\textbf
  {\bibinfo {volume} {4}},\ \bibinfo {pages} {706} (\bibinfo {year}
  {2008})}\BibitemShut {NoStop}%
\bibitem [{\citenamefont {Karzig}\ \emph {et~al.}(2015)\citenamefont {Karzig},
  \citenamefont {Bardyn}, \citenamefont {Lindner},\ and\ \citenamefont
  {Refael}}]{Karzig_topological}%
  \BibitemOpen
  \bibfield  {author} {\bibinfo {author} {\bibfnamefont {T.}~\bibnamefont
  {Karzig}}, \bibinfo {author} {\bibfnamefont {C.-E.}\ \bibnamefont {Bardyn}},
  \bibinfo {author} {\bibfnamefont {N.~H.}\ \bibnamefont {Lindner}}, \ and\
  \bibinfo {author} {\bibfnamefont {G.}~\bibnamefont {Refael}},\ }\href@noop {}
  {\bibfield  {journal} {\bibinfo  {journal} {Physical Review X}\ }\textbf
  {\bibinfo {volume} {5}},\ \bibinfo {pages} {031001} (\bibinfo {year}
  {2015})}\BibitemShut {NoStop}%
\bibitem [{\citenamefont {Nalitov}\ \emph {et~al.}(2015)\citenamefont
  {Nalitov}, \citenamefont {Solnyshkov},\ and\ \citenamefont
  {Malpuech}}]{Nalitov_topological}%
  \BibitemOpen
  \bibfield  {author} {\bibinfo {author} {\bibfnamefont {A.}~\bibnamefont
  {Nalitov}}, \bibinfo {author} {\bibfnamefont {D.}~\bibnamefont {Solnyshkov}},
  \ and\ \bibinfo {author} {\bibfnamefont {G.}~\bibnamefont {Malpuech}},\
  }\href@noop {} {\bibfield  {journal} {\bibinfo  {journal} {Physical review
  letters}\ }\textbf {\bibinfo {volume} {114}},\ \bibinfo {pages} {116401}
  (\bibinfo {year} {2015})}\BibitemShut {NoStop}%
\bibitem [{\citenamefont {Lubatsch}\ and\ \citenamefont
  {Frank}(2019)}]{lubatsch2019behavior}%
  \BibitemOpen
  \bibfield  {author} {\bibinfo {author} {\bibfnamefont {A.}~\bibnamefont
  {Lubatsch}}\ and\ \bibinfo {author} {\bibfnamefont {R.}~\bibnamefont
  {Frank}},\ }\href@noop {} {\bibfield  {journal} {\bibinfo  {journal}
  {Symmetry}\ }\textbf {\bibinfo {volume} {11}},\ \bibinfo {pages} {1246}
  (\bibinfo {year} {2019})}\BibitemShut {NoStop}%
\bibitem [{\citenamefont {Wouters}\ \emph {et~al.}(2008)\citenamefont
  {Wouters}, \citenamefont {Carusotto},\ and\ \citenamefont
  {Ciuti}}]{Wouters_nonequilibrium}%
  \BibitemOpen
  \bibfield  {author} {\bibinfo {author} {\bibfnamefont {M.}~\bibnamefont
  {Wouters}}, \bibinfo {author} {\bibfnamefont {I.}~\bibnamefont {Carusotto}},
  \ and\ \bibinfo {author} {\bibfnamefont {C.}~\bibnamefont {Ciuti}},\
  }\href@noop {} {\bibfield  {journal} {\bibinfo  {journal} {Physical Review
  B}\ }\textbf {\bibinfo {volume} {77}},\ \bibinfo {pages} {115340} (\bibinfo
  {year} {2008})}\BibitemShut {NoStop}%
\bibitem [{\citenamefont {Szyma{\'n}ska}\ \emph {et~al.}(2006)\citenamefont
  {Szyma{\'n}ska}, \citenamefont {Keeling},\ and\ \citenamefont
  {Littlewood}}]{Szymanska_nonequilibrium}%
  \BibitemOpen
  \bibfield  {author} {\bibinfo {author} {\bibfnamefont {M.}~\bibnamefont
  {Szyma{\'n}ska}}, \bibinfo {author} {\bibfnamefont {J.}~\bibnamefont
  {Keeling}}, \ and\ \bibinfo {author} {\bibfnamefont {P.}~\bibnamefont
  {Littlewood}},\ }\href@noop {} {\bibfield  {journal} {\bibinfo  {journal}
  {Physical review letters}\ }\textbf {\bibinfo {volume} {96}},\ \bibinfo
  {pages} {230602} (\bibinfo {year} {2006})}\BibitemShut {NoStop}%
\bibitem [{\citenamefont {De~Giorgi}\ \emph {et~al.}(2012)\citenamefont
  {De~Giorgi}, \citenamefont {Ballarini}, \citenamefont {Cancellieri},
  \citenamefont {Marchetti}, \citenamefont {Szymanska}, \citenamefont
  {Tejedor}, \citenamefont {Cingolani}, \citenamefont {Giacobino},
  \citenamefont {Bramati}, \citenamefont {Gigli} \emph
  {et~al.}}]{de2012control}%
  \BibitemOpen
  \bibfield  {author} {\bibinfo {author} {\bibfnamefont {M.}~\bibnamefont
  {De~Giorgi}}, \bibinfo {author} {\bibfnamefont {D.}~\bibnamefont
  {Ballarini}}, \bibinfo {author} {\bibfnamefont {E.}~\bibnamefont
  {Cancellieri}}, \bibinfo {author} {\bibfnamefont {F.}~\bibnamefont
  {Marchetti}}, \bibinfo {author} {\bibfnamefont {M.}~\bibnamefont
  {Szymanska}}, \bibinfo {author} {\bibfnamefont {C.}~\bibnamefont {Tejedor}},
  \bibinfo {author} {\bibfnamefont {R.}~\bibnamefont {Cingolani}}, \bibinfo
  {author} {\bibfnamefont {E.}~\bibnamefont {Giacobino}}, \bibinfo {author}
  {\bibfnamefont {A.}~\bibnamefont {Bramati}}, \bibinfo {author} {\bibfnamefont
  {G.}~\bibnamefont {Gigli}},  \emph {et~al.},\ }\href@noop {} {\bibfield
  {journal} {\bibinfo  {journal} {Physical review letters}\ }\textbf {\bibinfo
  {volume} {109}},\ \bibinfo {pages} {266407} (\bibinfo {year}
  {2012})}\BibitemShut {NoStop}%
\bibitem [{\citenamefont {Marsault}\ \emph {et~al.}(2015)\citenamefont
  {Marsault}, \citenamefont {Nguyen}, \citenamefont {Tanese}, \citenamefont
  {Lema{\^\i}tre}, \citenamefont {Galopin}, \citenamefont {Sagnes},
  \citenamefont {Amo},\ and\ \citenamefont {Bloch}}]{marsault2015realization}%
  \BibitemOpen
  \bibfield  {author} {\bibinfo {author} {\bibfnamefont {F.}~\bibnamefont
  {Marsault}}, \bibinfo {author} {\bibfnamefont {H.~S.}\ \bibnamefont
  {Nguyen}}, \bibinfo {author} {\bibfnamefont {D.}~\bibnamefont {Tanese}},
  \bibinfo {author} {\bibfnamefont {A.}~\bibnamefont {Lema{\^\i}tre}}, \bibinfo
  {author} {\bibfnamefont {E.}~\bibnamefont {Galopin}}, \bibinfo {author}
  {\bibfnamefont {I.}~\bibnamefont {Sagnes}}, \bibinfo {author} {\bibfnamefont
  {A.}~\bibnamefont {Amo}}, \ and\ \bibinfo {author} {\bibfnamefont
  {J.}~\bibnamefont {Bloch}},\ }\href@noop {} {\bibfield  {journal} {\bibinfo
  {journal} {Applied Physics Letters}\ }\textbf {\bibinfo {volume} {107}},\
  \bibinfo {pages} {201115} (\bibinfo {year} {2015})}\BibitemShut {NoStop}%
\bibitem [{\citenamefont {Berloff}\ \emph {et~al.}(2017)\citenamefont
  {Berloff}, \citenamefont {Silva}, \citenamefont {Kalinin}, \citenamefont
  {Askitopoulos}, \citenamefont {T{\"o}pfer}, \citenamefont {Cilibrizzi},
  \citenamefont {Langbein},\ and\ \citenamefont
  {Lagoudakis}}]{berloff2017realizing}%
  \BibitemOpen
  \bibfield  {author} {\bibinfo {author} {\bibfnamefont {N.~G.}\ \bibnamefont
  {Berloff}}, \bibinfo {author} {\bibfnamefont {M.}~\bibnamefont {Silva}},
  \bibinfo {author} {\bibfnamefont {K.}~\bibnamefont {Kalinin}}, \bibinfo
  {author} {\bibfnamefont {A.}~\bibnamefont {Askitopoulos}}, \bibinfo {author}
  {\bibfnamefont {J.~D.}\ \bibnamefont {T{\"o}pfer}}, \bibinfo {author}
  {\bibfnamefont {P.}~\bibnamefont {Cilibrizzi}}, \bibinfo {author}
  {\bibfnamefont {W.}~\bibnamefont {Langbein}}, \ and\ \bibinfo {author}
  {\bibfnamefont {P.~G.}\ \bibnamefont {Lagoudakis}},\ }\href@noop {}
  {\bibfield  {journal} {\bibinfo  {journal} {Nature materials}\ }\textbf
  {\bibinfo {volume} {16}},\ \bibinfo {pages} {1120} (\bibinfo {year}
  {2017})}\BibitemShut {NoStop}%
\bibitem [{\citenamefont {Gao}\ \emph {et~al.}(2012)\citenamefont {Gao},
  \citenamefont {Eldridge}, \citenamefont {Liew}, \citenamefont {Tsintzos},
  \citenamefont {Stavrinidis}, \citenamefont {Deligeorgis}, \citenamefont
  {Hatzopoulos},\ and\ \citenamefont {Savvidis}}]{Gao_transistor}%
  \BibitemOpen
  \bibfield  {author} {\bibinfo {author} {\bibfnamefont {T.}~\bibnamefont
  {Gao}}, \bibinfo {author} {\bibfnamefont {P.}~\bibnamefont {Eldridge}},
  \bibinfo {author} {\bibfnamefont {T.~C.~H.}\ \bibnamefont {Liew}}, \bibinfo
  {author} {\bibfnamefont {S.}~\bibnamefont {Tsintzos}}, \bibinfo {author}
  {\bibfnamefont {G.}~\bibnamefont {Stavrinidis}}, \bibinfo {author}
  {\bibfnamefont {G.}~\bibnamefont {Deligeorgis}}, \bibinfo {author}
  {\bibfnamefont {Z.}~\bibnamefont {Hatzopoulos}}, \ and\ \bibinfo {author}
  {\bibfnamefont {P.}~\bibnamefont {Savvidis}},\ }\href@noop {} {\bibfield
  {journal} {\bibinfo  {journal} {Physical Review B}\ }\textbf {\bibinfo
  {volume} {85}},\ \bibinfo {pages} {235102} (\bibinfo {year}
  {2012})}\BibitemShut {NoStop}%
\bibitem [{\citenamefont {Ballarini}\ \emph {et~al.}(2013)\citenamefont
  {Ballarini}, \citenamefont {De~Giorgi}, \citenamefont {Cancellieri},
  \citenamefont {Houdr{\'e}}, \citenamefont {Giacobino}, \citenamefont
  {Cingolani}, \citenamefont {Bramati}, \citenamefont {Gigli},\ and\
  \citenamefont {Sanvitto}}]{Ballarini_transistor}%
  \BibitemOpen
  \bibfield  {author} {\bibinfo {author} {\bibfnamefont {D.}~\bibnamefont
  {Ballarini}}, \bibinfo {author} {\bibfnamefont {M.}~\bibnamefont
  {De~Giorgi}}, \bibinfo {author} {\bibfnamefont {E.}~\bibnamefont
  {Cancellieri}}, \bibinfo {author} {\bibfnamefont {R.}~\bibnamefont
  {Houdr{\'e}}}, \bibinfo {author} {\bibfnamefont {E.}~\bibnamefont
  {Giacobino}}, \bibinfo {author} {\bibfnamefont {R.}~\bibnamefont
  {Cingolani}}, \bibinfo {author} {\bibfnamefont {A.}~\bibnamefont {Bramati}},
  \bibinfo {author} {\bibfnamefont {G.}~\bibnamefont {Gigli}}, \ and\ \bibinfo
  {author} {\bibfnamefont {D.}~\bibnamefont {Sanvitto}},\ }\href@noop {}
  {\bibfield  {journal} {\bibinfo  {journal} {Nature communications}\ }\textbf
  {\bibinfo {volume} {4}},\ \bibinfo {pages} {1778} (\bibinfo {year}
  {2013})}\BibitemShut {NoStop}%
\bibitem [{\citenamefont {Zasedatelev}\ \emph {et~al.}(2019)\citenamefont
  {Zasedatelev}, \citenamefont {Baranikov}, \citenamefont {Urbonas},
  \citenamefont {Scafirimuto}, \citenamefont {Scherf}, \citenamefont
  {St{\"o}ferle}, \citenamefont {Mahrt},\ and\ \citenamefont
  {Lagoudakis}}]{zasedatelev2019room}%
  \BibitemOpen
  \bibfield  {author} {\bibinfo {author} {\bibfnamefont {A.~V.}\ \bibnamefont
  {Zasedatelev}}, \bibinfo {author} {\bibfnamefont {A.~V.}\ \bibnamefont
  {Baranikov}}, \bibinfo {author} {\bibfnamefont {D.}~\bibnamefont {Urbonas}},
  \bibinfo {author} {\bibfnamefont {F.}~\bibnamefont {Scafirimuto}}, \bibinfo
  {author} {\bibfnamefont {U.}~\bibnamefont {Scherf}}, \bibinfo {author}
  {\bibfnamefont {T.}~\bibnamefont {St{\"o}ferle}}, \bibinfo {author}
  {\bibfnamefont {R.~F.}\ \bibnamefont {Mahrt}}, \ and\ \bibinfo {author}
  {\bibfnamefont {P.~G.}\ \bibnamefont {Lagoudakis}},\ }\href@noop {}
  {\bibfield  {journal} {\bibinfo  {journal} {Nature Photonics}\ }\textbf
  {\bibinfo {volume} {13}},\ \bibinfo {pages} {378} (\bibinfo {year}
  {2019})}\BibitemShut {NoStop}%
\bibitem [{\citenamefont {Amo}\ \emph {et~al.}(2010)\citenamefont {Amo},
  \citenamefont {Liew}, \citenamefont {Adrados}, \citenamefont {Houdr{\'e}},
  \citenamefont {Giacobino}, \citenamefont {Kavokin},\ and\ \citenamefont
  {Bramati}}]{Amo_switch}%
  \BibitemOpen
  \bibfield  {author} {\bibinfo {author} {\bibfnamefont {A.}~\bibnamefont
  {Amo}}, \bibinfo {author} {\bibfnamefont {T.}~\bibnamefont {Liew}}, \bibinfo
  {author} {\bibfnamefont {C.}~\bibnamefont {Adrados}}, \bibinfo {author}
  {\bibfnamefont {R.}~\bibnamefont {Houdr{\'e}}}, \bibinfo {author}
  {\bibfnamefont {E.}~\bibnamefont {Giacobino}}, \bibinfo {author}
  {\bibfnamefont {A.}~\bibnamefont {Kavokin}}, \ and\ \bibinfo {author}
  {\bibfnamefont {A.}~\bibnamefont {Bramati}},\ }\href@noop {} {\bibfield
  {journal} {\bibinfo  {journal} {Nature Photonics}\ }\textbf {\bibinfo
  {volume} {4}},\ \bibinfo {pages} {361} (\bibinfo {year} {2010})}\BibitemShut
  {NoStop}%
\bibitem [{\citenamefont {Frank}(2012)}]{frank2012coherent}%
  \BibitemOpen
  \bibfield  {author} {\bibinfo {author} {\bibfnamefont {R.}~\bibnamefont
  {Frank}},\ }\href@noop {} {\bibfield  {journal} {\bibinfo  {journal}
  {Physical Review B}\ }\textbf {\bibinfo {volume} {85}},\ \bibinfo {pages}
  {195463} (\bibinfo {year} {2012})}\BibitemShut {NoStop}%
\bibitem [{\citenamefont {Frank}(2013)}]{frank2013non}%
  \BibitemOpen
  \bibfield  {author} {\bibinfo {author} {\bibfnamefont {R.}~\bibnamefont
  {Frank}},\ }\href@noop {} {\bibfield  {journal} {\bibinfo  {journal} {Annalen
  der Physik}\ }\textbf {\bibinfo {volume} {525}},\ \bibinfo {pages} {66}
  (\bibinfo {year} {2013})}\BibitemShut {NoStop}%
\bibitem [{\citenamefont {Ozawa}\ \emph {et~al.}(2019)\citenamefont {Ozawa},
  \citenamefont {Price}, \citenamefont {Amo}, \citenamefont {Goldman},
  \citenamefont {Hafezi}, \citenamefont {Lu}, \citenamefont {Rechtsman},
  \citenamefont {Schuster}, \citenamefont {Simon}, \citenamefont {Zilberberg}
  \emph {et~al.}}]{ozawa2019topological}%
  \BibitemOpen
  \bibfield  {author} {\bibinfo {author} {\bibfnamefont {T.}~\bibnamefont
  {Ozawa}}, \bibinfo {author} {\bibfnamefont {H.~M.}\ \bibnamefont {Price}},
  \bibinfo {author} {\bibfnamefont {A.}~\bibnamefont {Amo}}, \bibinfo {author}
  {\bibfnamefont {N.}~\bibnamefont {Goldman}}, \bibinfo {author} {\bibfnamefont
  {M.}~\bibnamefont {Hafezi}}, \bibinfo {author} {\bibfnamefont
  {L.}~\bibnamefont {Lu}}, \bibinfo {author} {\bibfnamefont {M.~C.}\
  \bibnamefont {Rechtsman}}, \bibinfo {author} {\bibfnamefont {D.}~\bibnamefont
  {Schuster}}, \bibinfo {author} {\bibfnamefont {J.}~\bibnamefont {Simon}},
  \bibinfo {author} {\bibfnamefont {O.}~\bibnamefont {Zilberberg}},  \emph
  {et~al.},\ }\href@noop {} {\bibfield  {journal} {\bibinfo  {journal} {Reviews
  of Modern Physics}\ }\textbf {\bibinfo {volume} {91}},\ \bibinfo {pages}
  {015006} (\bibinfo {year} {2019})}\BibitemShut {NoStop}%
\bibitem [{\citenamefont {Kalinin}\ and\ \citenamefont
  {Berloff}(2020)}]{kalinin2019towards}%
  \BibitemOpen
  \bibfield  {author} {\bibinfo {author} {\bibfnamefont {K.~P.}\ \bibnamefont
  {Kalinin}}\ and\ \bibinfo {author} {\bibfnamefont {N.~G.}\ \bibnamefont
  {Berloff}},\ }\href@noop {} {\bibfield  {journal} {\bibinfo  {journal}
  {Advanced Quantum Technologies}\ }\textbf {\bibinfo {volume} {3}},\ \bibinfo
  {pages} {1900065} (\bibinfo {year} {2020})}\BibitemShut {NoStop}%
\bibitem [{\citenamefont {Na}\ and\ \citenamefont
  {Yamamoto}(2010)}]{na2010massive}%
  \BibitemOpen
  \bibfield  {author} {\bibinfo {author} {\bibfnamefont {N.}~\bibnamefont
  {Na}}\ and\ \bibinfo {author} {\bibfnamefont {Y.}~\bibnamefont {Yamamoto}},\
  }\href@noop {} {\bibfield  {journal} {\bibinfo  {journal} {New Journal of
  Physics}\ }\textbf {\bibinfo {volume} {12}},\ \bibinfo {pages} {123001}
  (\bibinfo {year} {2010})}\BibitemShut {NoStop}%
\bibitem [{\citenamefont {Tanese}\ \emph {et~al.}(2013)\citenamefont {Tanese},
  \citenamefont {Flayac}, \citenamefont {Solnyshkov}, \citenamefont {Amo},
  \citenamefont {Lemaitre}, \citenamefont {Galopin}, \citenamefont {Braive},
  \citenamefont {Senellart}, \citenamefont {Sagnes}, \citenamefont {Malpuech}
  \emph {et~al.}}]{tanese2013polariton}%
  \BibitemOpen
  \bibfield  {author} {\bibinfo {author} {\bibfnamefont {D.}~\bibnamefont
  {Tanese}}, \bibinfo {author} {\bibfnamefont {H.}~\bibnamefont {Flayac}},
  \bibinfo {author} {\bibfnamefont {D.}~\bibnamefont {Solnyshkov}}, \bibinfo
  {author} {\bibfnamefont {A.}~\bibnamefont {Amo}}, \bibinfo {author}
  {\bibfnamefont {A.}~\bibnamefont {Lemaitre}}, \bibinfo {author}
  {\bibfnamefont {E.}~\bibnamefont {Galopin}}, \bibinfo {author} {\bibfnamefont
  {R.}~\bibnamefont {Braive}}, \bibinfo {author} {\bibfnamefont
  {P.}~\bibnamefont {Senellart}}, \bibinfo {author} {\bibfnamefont
  {I.}~\bibnamefont {Sagnes}}, \bibinfo {author} {\bibfnamefont
  {G.}~\bibnamefont {Malpuech}},  \emph {et~al.},\ }\href@noop {} {\bibfield
  {journal} {\bibinfo  {journal} {Nature communications}\ }\textbf {\bibinfo
  {volume} {4}},\ \bibinfo {pages} {1} (\bibinfo {year} {2013})}\BibitemShut
  {NoStop}%
\bibitem [{\citenamefont {St-Jean}\ \emph {et~al.}(2017)\citenamefont
  {St-Jean}, \citenamefont {Goblot}, \citenamefont {Galopin}, \citenamefont
  {Lema{\^\i}tre}, \citenamefont {Ozawa}, \citenamefont {Le~Gratiet},
  \citenamefont {Sagnes}, \citenamefont {Bloch},\ and\ \citenamefont
  {Amo}}]{st2017lasing}%
  \BibitemOpen
  \bibfield  {author} {\bibinfo {author} {\bibfnamefont {P.}~\bibnamefont
  {St-Jean}}, \bibinfo {author} {\bibfnamefont {V.}~\bibnamefont {Goblot}},
  \bibinfo {author} {\bibfnamefont {E.}~\bibnamefont {Galopin}}, \bibinfo
  {author} {\bibfnamefont {A.}~\bibnamefont {Lema{\^\i}tre}}, \bibinfo {author}
  {\bibfnamefont {T.}~\bibnamefont {Ozawa}}, \bibinfo {author} {\bibfnamefont
  {L.}~\bibnamefont {Le~Gratiet}}, \bibinfo {author} {\bibfnamefont
  {I.}~\bibnamefont {Sagnes}}, \bibinfo {author} {\bibfnamefont
  {J.}~\bibnamefont {Bloch}}, \ and\ \bibinfo {author} {\bibfnamefont
  {A.}~\bibnamefont {Amo}},\ }\href@noop {} {\bibfield  {journal} {\bibinfo
  {journal} {Nature Photonics}\ }\textbf {\bibinfo {volume} {11}},\ \bibinfo
  {pages} {651} (\bibinfo {year} {2017})}\BibitemShut {NoStop}%
\bibitem [{\citenamefont {Mittal}\ \emph {et~al.}(2014)\citenamefont {Mittal},
  \citenamefont {Fan}, \citenamefont {Faez}, \citenamefont {Migdall},
  \citenamefont {Taylor},\ and\ \citenamefont
  {Hafezi}}]{mittal2014topologically}%
  \BibitemOpen
  \bibfield  {author} {\bibinfo {author} {\bibfnamefont {S.}~\bibnamefont
  {Mittal}}, \bibinfo {author} {\bibfnamefont {J.}~\bibnamefont {Fan}},
  \bibinfo {author} {\bibfnamefont {S.}~\bibnamefont {Faez}}, \bibinfo {author}
  {\bibfnamefont {A.}~\bibnamefont {Migdall}}, \bibinfo {author} {\bibfnamefont
  {J.}~\bibnamefont {Taylor}}, \ and\ \bibinfo {author} {\bibfnamefont
  {M.}~\bibnamefont {Hafezi}},\ }\href@noop {} {\bibfield  {journal} {\bibinfo
  {journal} {Physical review letters}\ }\textbf {\bibinfo {volume} {113}},\
  \bibinfo {pages} {087403} (\bibinfo {year} {2014})}\BibitemShut {NoStop}%
\bibitem [{\citenamefont {Su}\ \emph {et~al.}(2018)\citenamefont {Su},
  \citenamefont {Wang}, \citenamefont {Zhao}, \citenamefont {Xing},
  \citenamefont {Zhao}, \citenamefont {Diederichs}, \citenamefont {Liew},\ and\
  \citenamefont {Xiong}}]{su2018room}%
  \BibitemOpen
  \bibfield  {author} {\bibinfo {author} {\bibfnamefont {R.}~\bibnamefont
  {Su}}, \bibinfo {author} {\bibfnamefont {J.}~\bibnamefont {Wang}}, \bibinfo
  {author} {\bibfnamefont {J.}~\bibnamefont {Zhao}}, \bibinfo {author}
  {\bibfnamefont {J.}~\bibnamefont {Xing}}, \bibinfo {author} {\bibfnamefont
  {W.}~\bibnamefont {Zhao}}, \bibinfo {author} {\bibfnamefont {C.}~\bibnamefont
  {Diederichs}}, \bibinfo {author} {\bibfnamefont {T.~C.}\ \bibnamefont
  {Liew}}, \ and\ \bibinfo {author} {\bibfnamefont {Q.}~\bibnamefont {Xiong}},\
  }\href@noop {} {\bibfield  {journal} {\bibinfo  {journal} {Science advances}\
  }\textbf {\bibinfo {volume} {4}},\ \bibinfo {pages} {eaau0244} (\bibinfo
  {year} {2018})}\BibitemShut {NoStop}%
\bibitem [{\citenamefont {Su}\ \emph {et~al.}(2019)\citenamefont {Su},
  \citenamefont {Ghosh}, \citenamefont {Liu}, \citenamefont {Diederichs},
  \citenamefont {Liew},\ and\ \citenamefont {Xiong}}]{su2019observation}%
  \BibitemOpen
  \bibfield  {author} {\bibinfo {author} {\bibfnamefont {R.}~\bibnamefont
  {Su}}, \bibinfo {author} {\bibfnamefont {S.}~\bibnamefont {Ghosh}}, \bibinfo
  {author} {\bibfnamefont {S.}~\bibnamefont {Liu}}, \bibinfo {author}
  {\bibfnamefont {C.}~\bibnamefont {Diederichs}}, \bibinfo {author}
  {\bibfnamefont {T.~C.}\ \bibnamefont {Liew}}, \ and\ \bibinfo {author}
  {\bibfnamefont {Q.}~\bibnamefont {Xiong}},\ }\href@noop {} {\bibfield
  {journal} {\bibinfo  {journal} {arXiv preprint arXiv:1906.11566}\ } (\bibinfo
  {year} {2019})}\BibitemShut {NoStop}%
\bibitem [{\citenamefont {Schneider}\ \emph {et~al.}(2016)\citenamefont
  {Schneider}, \citenamefont {Winkler}, \citenamefont {Fraser}, \citenamefont
  {Kamp}, \citenamefont {Yamamoto}, \citenamefont {Ostrovskaya},\ and\
  \citenamefont {H{\"o}fling}}]{Schneider_landscape}%
  \BibitemOpen
  \bibfield  {author} {\bibinfo {author} {\bibfnamefont {C.}~\bibnamefont
  {Schneider}}, \bibinfo {author} {\bibfnamefont {K.}~\bibnamefont {Winkler}},
  \bibinfo {author} {\bibfnamefont {M.}~\bibnamefont {Fraser}}, \bibinfo
  {author} {\bibfnamefont {M.}~\bibnamefont {Kamp}}, \bibinfo {author}
  {\bibfnamefont {Y.}~\bibnamefont {Yamamoto}}, \bibinfo {author}
  {\bibfnamefont {E.}~\bibnamefont {Ostrovskaya}}, \ and\ \bibinfo {author}
  {\bibfnamefont {S.}~\bibnamefont {H{\"o}fling}},\ }\href@noop {} {\bibfield
  {journal} {\bibinfo  {journal} {Reports on Progress in Physics}\ }\textbf
  {\bibinfo {volume} {80}},\ \bibinfo {pages} {016503} (\bibinfo {year}
  {2016})}\BibitemShut {NoStop}%
\bibitem [{\citenamefont {Sanvitto}\ and\ \citenamefont
  {K{\'e}na-Cohen}(2016)}]{sanvitto2016road}%
  \BibitemOpen
  \bibfield  {author} {\bibinfo {author} {\bibfnamefont {D.}~\bibnamefont
  {Sanvitto}}\ and\ \bibinfo {author} {\bibfnamefont {S.}~\bibnamefont
  {K{\'e}na-Cohen}},\ }\href@noop {} {\bibfield  {journal} {\bibinfo  {journal}
  {Nature materials}\ }\textbf {\bibinfo {volume} {15}},\ \bibinfo {pages}
  {1061} (\bibinfo {year} {2016})}\BibitemShut {NoStop}%
\bibitem [{\citenamefont {Rajendran}\ \emph {et~al.}(2019)\citenamefont
  {Rajendran}, \citenamefont {Wei}, \citenamefont {Ohadi}, \citenamefont
  {Ruseckas}, \citenamefont {Turnbull},\ and\ \citenamefont
  {Samuel}}]{rajendran2019low}%
  \BibitemOpen
  \bibfield  {author} {\bibinfo {author} {\bibfnamefont {S.~K.}\ \bibnamefont
  {Rajendran}}, \bibinfo {author} {\bibfnamefont {M.}~\bibnamefont {Wei}},
  \bibinfo {author} {\bibfnamefont {H.}~\bibnamefont {Ohadi}}, \bibinfo
  {author} {\bibfnamefont {A.}~\bibnamefont {Ruseckas}}, \bibinfo {author}
  {\bibfnamefont {G.~A.}\ \bibnamefont {Turnbull}}, \ and\ \bibinfo {author}
  {\bibfnamefont {I.~D.}\ \bibnamefont {Samuel}},\ }\href@noop {} {\bibfield
  {journal} {\bibinfo  {journal} {Advanced Optical Materials}\ ,\ \bibinfo
  {pages} {1801791}} (\bibinfo {year} {2019})}\BibitemShut {NoStop}%
\bibitem [{\citenamefont {Keeling}\ and\ \citenamefont
  {Berloff}(2008)}]{keeling2008spontaneous}%
  \BibitemOpen
  \bibfield  {author} {\bibinfo {author} {\bibfnamefont {J.}~\bibnamefont
  {Keeling}}\ and\ \bibinfo {author} {\bibfnamefont {N.~G.}\ \bibnamefont
  {Berloff}},\ }\href@noop {} {\bibfield  {journal} {\bibinfo  {journal}
  {Physical review letters}\ }\textbf {\bibinfo {volume} {100}},\ \bibinfo
  {pages} {250401} (\bibinfo {year} {2008})}\BibitemShut {NoStop}%
\bibitem [{\citenamefont {Newell}\ \emph {et~al.}(1993)\citenamefont {Newell},
  \citenamefont {Passot},\ and\ \citenamefont {Lega}}]{Newell_order}%
  \BibitemOpen
  \bibfield  {author} {\bibinfo {author} {\bibfnamefont {A.~C.}\ \bibnamefont
  {Newell}}, \bibinfo {author} {\bibfnamefont {T.}~\bibnamefont {Passot}}, \
  and\ \bibinfo {author} {\bibfnamefont {J.}~\bibnamefont {Lega}},\ }\href@noop
  {} {\bibfield  {journal} {\bibinfo  {journal} {Annual review of fluid
  mechanics}\ }\textbf {\bibinfo {volume} {25}},\ \bibinfo {pages} {399}
  (\bibinfo {year} {1993})}\BibitemShut {NoStop}%
\bibitem [{\citenamefont {Pismen}\ and\ \citenamefont
  {Pismen}(1999)}]{Pismen_order}%
  \BibitemOpen
  \bibfield  {author} {\bibinfo {author} {\bibfnamefont {L.~M.}\ \bibnamefont
  {Pismen}}\ and\ \bibinfo {author} {\bibfnamefont {L.~M.}\ \bibnamefont
  {Pismen}},\ }\href@noop {} {\emph {\bibinfo {title} {Vortices in nonlinear
  fields: from liquid crystals to superfluids, from non-equilibrium patterns to
  cosmic strings}}},\ Vol.\ \bibinfo {volume} {100}\ (\bibinfo  {publisher}
  {Oxford University Press},\ \bibinfo {year} {1999})\BibitemShut {NoStop}%
\bibitem [{\citenamefont {Cross}\ and\ \citenamefont
  {Hohenberg}(1993)}]{Cross_order}%
  \BibitemOpen
  \bibfield  {author} {\bibinfo {author} {\bibfnamefont {M.~C.}\ \bibnamefont
  {Cross}}\ and\ \bibinfo {author} {\bibfnamefont {P.~C.}\ \bibnamefont
  {Hohenberg}},\ }\href@noop {} {\bibfield  {journal} {\bibinfo  {journal}
  {Reviews of modern physics}\ }\textbf {\bibinfo {volume} {65}},\ \bibinfo
  {pages} {851} (\bibinfo {year} {1993})}\BibitemShut {NoStop}%
\bibitem [{\citenamefont {Cristofolini}\ \emph {et~al.}(2013)\citenamefont
  {Cristofolini}, \citenamefont {Dreismann}, \citenamefont {Christmann},
  \citenamefont {Franchetti}, \citenamefont {Berloff}, \citenamefont {Tsotsis},
  \citenamefont {Hatzopoulos}, \citenamefont {Savvidis},\ and\ \citenamefont
  {Baumberg}}]{cristofolini2013optical}%
  \BibitemOpen
  \bibfield  {author} {\bibinfo {author} {\bibfnamefont {P.}~\bibnamefont
  {Cristofolini}}, \bibinfo {author} {\bibfnamefont {A.}~\bibnamefont
  {Dreismann}}, \bibinfo {author} {\bibfnamefont {G.}~\bibnamefont
  {Christmann}}, \bibinfo {author} {\bibfnamefont {G.}~\bibnamefont
  {Franchetti}}, \bibinfo {author} {\bibfnamefont {N.}~\bibnamefont {Berloff}},
  \bibinfo {author} {\bibfnamefont {P.}~\bibnamefont {Tsotsis}}, \bibinfo
  {author} {\bibfnamefont {Z.}~\bibnamefont {Hatzopoulos}}, \bibinfo {author}
  {\bibfnamefont {P.}~\bibnamefont {Savvidis}}, \ and\ \bibinfo {author}
  {\bibfnamefont {J.}~\bibnamefont {Baumberg}},\ }\href@noop {} {\bibfield
  {journal} {\bibinfo  {journal} {Physical review letters}\ }\textbf {\bibinfo
  {volume} {110}},\ \bibinfo {pages} {186403} (\bibinfo {year}
  {2013})}\BibitemShut {NoStop}%
\bibitem [{\citenamefont {Bronski}\ \emph
  {et~al.}(2001{\natexlab{a}})\citenamefont {Bronski}, \citenamefont {Carr},
  \citenamefont {Carretero-Gonz{\'a}lez}, \citenamefont {Deconinck},
  \citenamefont {Kutz},\ and\ \citenamefont {Promislow}}]{Bronski_attractive}%
  \BibitemOpen
  \bibfield  {author} {\bibinfo {author} {\bibfnamefont {J.~C.}\ \bibnamefont
  {Bronski}}, \bibinfo {author} {\bibfnamefont {L.~D.}\ \bibnamefont {Carr}},
  \bibinfo {author} {\bibfnamefont {R.}~\bibnamefont {Carretero-Gonz{\'a}lez}},
  \bibinfo {author} {\bibfnamefont {B.}~\bibnamefont {Deconinck}}, \bibinfo
  {author} {\bibfnamefont {J.~N.}\ \bibnamefont {Kutz}}, \ and\ \bibinfo
  {author} {\bibfnamefont {K.}~\bibnamefont {Promislow}},\ }\href@noop {}
  {\bibfield  {journal} {\bibinfo  {journal} {Physical Review E}\ }\textbf
  {\bibinfo {volume} {64}},\ \bibinfo {pages} {056615} (\bibinfo {year}
  {2001}{\natexlab{a}})}\BibitemShut {NoStop}%
\bibitem [{\citenamefont {Bronski}\ \emph
  {et~al.}(2001{\natexlab{b}})\citenamefont {Bronski}, \citenamefont {Carr},
  \citenamefont {Deconinck}, \citenamefont {Kutz},\ and\ \citenamefont
  {Promislow}}]{Bronski_repulsive}%
  \BibitemOpen
  \bibfield  {author} {\bibinfo {author} {\bibfnamefont {J.~C.}\ \bibnamefont
  {Bronski}}, \bibinfo {author} {\bibfnamefont {L.~D.}\ \bibnamefont {Carr}},
  \bibinfo {author} {\bibfnamefont {B.}~\bibnamefont {Deconinck}}, \bibinfo
  {author} {\bibfnamefont {J.~N.}\ \bibnamefont {Kutz}}, \ and\ \bibinfo
  {author} {\bibfnamefont {K.}~\bibnamefont {Promislow}},\ }\href@noop {}
  {\bibfield  {journal} {\bibinfo  {journal} {Physical Review E}\ }\textbf
  {\bibinfo {volume} {63}},\ \bibinfo {pages} {036612} (\bibinfo {year}
  {2001}{\natexlab{b}})}\BibitemShut {NoStop}%
\bibitem [{\citenamefont {Alyatkin}\ \emph {et~al.}(2020)\citenamefont
  {Alyatkin}, \citenamefont {T{\"o}pfer}, \citenamefont {Askitopoulos},
  \citenamefont {Sigurdsson},\ and\ \citenamefont
  {Lagoudakis}}]{alyatkin2020optical}%
  \BibitemOpen
  \bibfield  {author} {\bibinfo {author} {\bibfnamefont {S.}~\bibnamefont
  {Alyatkin}}, \bibinfo {author} {\bibfnamefont {J.}~\bibnamefont
  {T{\"o}pfer}}, \bibinfo {author} {\bibfnamefont {A.}~\bibnamefont
  {Askitopoulos}}, \bibinfo {author} {\bibfnamefont {H.}~\bibnamefont
  {Sigurdsson}}, \ and\ \bibinfo {author} {\bibfnamefont {P.}~\bibnamefont
  {Lagoudakis}},\ }\href@noop {} {\bibfield  {journal} {\bibinfo  {journal}
  {Physical Review Letters}\ }\textbf {\bibinfo {volume} {124}},\ \bibinfo
  {pages} {207402} (\bibinfo {year} {2020})}\BibitemShut {NoStop}%
\bibitem [{\citenamefont {Estrecho}\ \emph {et~al.}(2019)\citenamefont
  {Estrecho}, \citenamefont {Gao}, \citenamefont {Bobrovska}, \citenamefont
  {Comber-Todd}, \citenamefont {Fraser}, \citenamefont {Steger}, \citenamefont
  {West}, \citenamefont {Pfeiffer}, \citenamefont {Levinsen}, \citenamefont
  {Parish} \emph {et~al.}}]{estrecho2019direct}%
  \BibitemOpen
  \bibfield  {author} {\bibinfo {author} {\bibfnamefont {E.}~\bibnamefont
  {Estrecho}}, \bibinfo {author} {\bibfnamefont {T.}~\bibnamefont {Gao}},
  \bibinfo {author} {\bibfnamefont {N.}~\bibnamefont {Bobrovska}}, \bibinfo
  {author} {\bibfnamefont {D.}~\bibnamefont {Comber-Todd}}, \bibinfo {author}
  {\bibfnamefont {M.}~\bibnamefont {Fraser}}, \bibinfo {author} {\bibfnamefont
  {M.}~\bibnamefont {Steger}}, \bibinfo {author} {\bibfnamefont
  {K.}~\bibnamefont {West}}, \bibinfo {author} {\bibfnamefont {L.}~\bibnamefont
  {Pfeiffer}}, \bibinfo {author} {\bibfnamefont {J.}~\bibnamefont {Levinsen}},
  \bibinfo {author} {\bibfnamefont {M.}~\bibnamefont {Parish}},  \emph
  {et~al.},\ }\href@noop {} {\bibfield  {journal} {\bibinfo  {journal}
  {Physical Review B}\ }\textbf {\bibinfo {volume} {100}},\ \bibinfo {pages}
  {035306} (\bibinfo {year} {2019})}\BibitemShut {NoStop}%
\bibitem [{\citenamefont {Lagoudakis}\ and\ \citenamefont
  {Berloff}(2017)}]{lagoudakis2017polariton}%
  \BibitemOpen
  \bibfield  {author} {\bibinfo {author} {\bibfnamefont {P.~G.}\ \bibnamefont
  {Lagoudakis}}\ and\ \bibinfo {author} {\bibfnamefont {N.~G.}\ \bibnamefont
  {Berloff}},\ }\href@noop {} {\bibfield  {journal} {\bibinfo  {journal} {New
  Journal of Physics}\ }\textbf {\bibinfo {volume} {19}},\ \bibinfo {pages}
  {125008} (\bibinfo {year} {2017})}\BibitemShut {NoStop}%
\end{thebibliography}%

\end{document}